\documentclass[letterpaper,pra,showpacs]{revtex4}
\usepackage{graphicx}
\usepackage{amsmath}
\usepackage{amssymb}

\begin{document}

\title{Criteria of quantum correlation in the measurement of continuous variables in optics}
\author{N. Treps, C. Fabre}
\affiliation{Laboratoire Kastler Brossel, UPMC, Case 74, 4 Place
Jussieu, 75252 Paris cedex 05, France}

\date{\today}

\begin{abstract}
The purpose of this short tutorial paper is to review various
criteria that have been used to characterize the quantum character
of correlations in optical systems, such as "gemellity", QND
correlation, intrication, EPR correlation and Bell correlation, to
discuss and compare them. This discussion, restricted to the case
of measurements of continuous optical variables, includes also an
extension of known criteria for "twin beams" to the case of
imbalanced correlations.
\end{abstract}


\maketitle

\section{Introduction}

One of the most striking features of quantum mechanics is the
existence of the so-called entangled states, i.e. of quantum
states $|\Psi>$ describing a system made of two separable parts
which cannot be written as a tensor product of quantum states
$|\Psi_1>$ and $|\Psi_2>$ describing separately each of the
subsystems~:
\begin{equation}
|\Psi> \neq|\psi_1> \otimes|\psi_2> \label{entangled}
\end{equation}
In such states, there exist strong correlations between
measurements performed on the sub-systems. These correlations have
been widely studied, almost from the onset of quantum physics, but
they still keep a part of their mystery, and therefore of their
attraction. The discovery that quantum correlations play an
irreplaceable role in information processing gave recently a new
impetus to their study.

The existence of correlations between different measurements is
obviously not a specific property of quantum physics~: it is
simply the consequence of a former interaction, whatever its
character, between the two parts submitted to the measurement.
Consequently, the observation or prediction of a correlation, even
perfect, between the measurements of two variables is not at all a
proof of the quantum character of the phenomenon under study, in
contrast to what can be found sometimes in articles. One can find
in the literature a great deal of criteria setting a border
between the classical and the quantum effects, differing by the
definitions of the quantum character of a given physical
situation. The purpose of the present paper is mainly tutorial :
it is to give a short overview of the different criteria which are
already well-known and extensively used in the literature, to
compare them and discuss their domain of relevance. We will also
introduce slight original additions to some already known
criteria, especially for the criterion of "gemellity" in the case
where the two correlated systems do not play symmetrical roles.

Of course, it is impossible to treat the problem of quantum
correlations in all its generality in such a short review. We will
restrict ourselves here to the domain of the so-called
\textit{continuous variables in optics}. More precisely we will
consider correlations between measurements performed on light
beams in the case where the photons cannot be distinguished
individually, and leave aside the correlations between
photo-counts and between measurements performed on other
observables than the ones arising from photodetection. One of the
important application of this work is to be able to precisely
assess the quantum aspects of the correlations appearing between
two points of the transverse plane of an optical image.

After a short paragraph devoted to the introduction of the problem
and of the notations, we will successively focus on the different
criteria of quantum correlations, involving either a single
correlation measurement, and assessing first the impossibility of
a classical description of the phenomenon, then the QND character
of the correlation, or involving the measurement of two
correlations on non-commuting observables, and assessing first the
intrication of the state describing the system, then its EPR
character, and finally its impossibility of description in terms
of local hidden variables.

\section{Position of the problem, notations}

Let us consider two modes of the electromagnetic field, labelled 1
and 2, that can be separated without introducing losses in the
system, and that are measured by detectors situated at different
locations : the modes can differ either by the frequency, the
direction of propagation, the polarization, the transverse shape,
or by several of these characteristics. $\hat{a}_1$ and
$\hat{a}_2$ are the corresponding annihilation operators, and
\begin{equation}
\hat{X}^+_{i}=\hat{a}_i+\hat{a}_i^{\dagger} \quad
\hat{X}^-_{i}=\frac{\hat{a}_i-\hat{a}_i^{\dagger}}{i} \quad
(i=1,2)
\end{equation}
the quadrature operators. The measurement performed on these modes
can be either a direct photodetection, which measures the
fluctuations of the amplitude quadrature component, parallel to
the mean field $\bar{E}_i$  in the Fresnel representation plane,
or a balanced homodyne detection, which measures any quadrature
component. We will call in a generic way $\hat{X}_{i}$ the
quadrature component which is measured, and $\delta \hat{X}_{i}$
its fluctuations.

>From the fluctuations measured on a single detector, one can
deduce the quantity :
\begin{equation}
F_{i}=\langle \delta \hat{X}_{i}^2 \rangle
\end{equation}
equal to 1 when the field is in a coherent state. $F_i$ is the
Fano factor \cite{Fano, Bachor} of the beam in the case of a
direct photodetection, and the quadrature noise normalized to shot
noise in a homodyne measurement.

The simultaneous recording of the fluctuations measured by the two
detectors allows us to determine the normalized correlation
coefficient :
\begin{equation}
C_{12}=\frac{\langle \delta X_{1} \delta X_{2} \rangle}{(F_1
F_2)^{1/2}}
\end{equation}
which varies between $-1$ (perfect anti-correlation) and $1$
(perfect correlation).

For the sake of simplicity, we will assume in the following, when
it turns out to be necessary, that the system under study has
correlations, and not anti-correlations, and therefore that
$C_{12}$ is positive. All our following discussion can be readily
extended to the case of anti-correlations by exchanging the role
of sum and differences between the quantities measured on the two
modes. One can also note that the study performed here can be
applied to the fluctuations of non-optical physical systems, as
soon as a protocol exists to transfer these fluctuations to an
optical field, as is done in the case of cold atoms and
entanglement between light and cold atoms \cite{Polzik, Aurelien}.

Let us consider now the case of an optical experiment which gives
an experimental value of $C_{12}$ close to 1. In which respect can
one claim the quantum nature of the observed correlations ? That
is the question that we will address in the following sections.

\section{Non classical character of the correlated beams : "Twin beams"}

One knows that most of the optical phenomena can be explained by
using the so-called \textit{semi-classical approach of the
light-matter interaction}, in which a quantized matter interacts
with the electro-magnetic field treated as a classical quantity,
possibly endowed with classical fluctuations. Even the
photo-electric effect, for which the photon was introduced by
Einstein \cite{Einstein} lies in this category, which also
includes all interference effects that are directly measured on
the intensity of the field. In this model, the fluctuations which
exist in the photo-detection signal are due to the random
character of the "quantum jump" occurring in the atom because of
its irradiation by the classical field. The minimum noise measured
on the photo-detector is the shot noise, or standard quantum noise
limit. As explained in many textbooks on Quantum Optics
\cite{Mandel, Walls,Bachor}, it was realized in the seventies that
there existed light states which gave rise to measurements that
could not be accounted for within the semi-classical approach.
These states are named "non-classical states", and are unveiled by
measurements involving either intensity correlations between two
photocurrents, or intensity fluctuations of a given photocurrent
around the mean. The observation of photon antibunching using
single photon states has been historically the first unambiguous
experimental situation\cite{Mandel1} where a classical description
of the field was not able to account for the observed results. Let
us note that Herbert Walther has played a major role in the study
of these effects, using the light emitted by a single trapped ion
\cite{Walther1, Walther2}. One can therefore define a first level
 of quantum correlation by the following statement :\\

\textbf{Quantum correlation, level 1 :} \textit{The correlation
measured in the system cannot be described by a semi-classical
model involving classical electromagnetic fields
having classical fluctuations.}\\

In the domain of continuous variables and intense beams to which
we restrict the present discussion, one shows that the situations
where one records quadrature fluctuations above the standard
quantum noise limit can be described using the semi-classical
approach with classical stochastic electro-magnetic fields
\cite{Bachor, Loudon}, and that quantum fluctuations below the
this limit are only produced by non-classical states. Squeezed
states are one example of non-classical states. The border between
the classical and quantum world corresponds to the situation where
all the beams used in the experiment are in coherent and vacuum
states (Fano factor of 1). Furthermore, it is easy to show that
\textit{the classical character of the field is preserved by
linear "passive" optical devices}, which involve only linear,
energy-preserving, optical elements like beamsplitters and free
propagation.

To ascertain whether the correlation between two given beams can
be described in a classical frame or not, the simplest way is
therefore to process the two beams by all possible linear passive
optical devices : if one is able to produce in such a way a beam
having fluctuations below the quantum noise limit, the correlation
will be termed as "non-classical". This procedure is easy to
implement if the two beams have the same frequency. If the two
beams have different frequencies $\omega_1$ and $\omega_2$, it is
more difficult but still possible, at least in principle : the
noise reduction will then be measured by a homodyne detection
scheme using a local oscillator at frequency
$(\omega_1+\omega_2)/2$, and will appear at a noise frequency
$\Omega =|\omega_1-\omega_2|/2$.

\subsection{Classical correlation}

Let us first consider the simplest way to produce correlated beams
by classical means : one inserts a $50\%$ beamsplitter in a given
classical beam, which is thus divided into two output beams having
a degree of correlation that can be simply calculated. Taking into
account the vacuum mode entering through the unused port of the
beamsplitter, one obtains the following value for the correlation
obtained by splitting an input classical field on a $50\%$
beamsplitter :
\begin{equation}
[C_{12}]_{class}=\frac{F_{in}-1}{F_{in}+1}
\end{equation}
where $F_{in}$ is the Fano factor of the input beam on the
beamsplitter, or equivalently by :
\begin{equation}
[C_{12}]_{class}=1-\frac{1}{F} \label{C12}
\end{equation}
where $F$ is the common value of the Fano factor of the two beams
at the output of the beamsplitter ($F=(F_{in}+1)/2$).

Let us note that $C_{12}$ tends to 1 when $F_{in}$ goes to
infinity, i.e. when the vacuum noise of the second input can be
neglected with respect to the proper noise of the input beam. A
very strong correlation is therefore not always the sign of a
quantum origin : it can be just the reverse, and due to the fact
that the quantum fluctuations can be neglected in the problem !
The normalized correlation factor $C_{12}$ is thus not the most
unambiguous way to appreciate the quantum character of a
correlation.

\subsection{"Gemellity"}

Let us now exploit the operational definition of the quantum
character of the correlation given at the beginning of this
section, which is to use a linear passive operation which
transforms the correlation into a sub-shot noise beam. In the case
of two beams, this operation simply consists of recombining the
beams on a beamsplitter of variable transmission and reflection
after variable optical paths. The phases are adjusted so that one
mixes the relevant quadrature components $X_{1}$ and $X_{2}$. One
eventually obtains a beam having quadrature fluctuations $\delta
\hat{X}_{out}$ given by :

\begin{equation}
\delta \hat{X}_{out}= r \delta \hat{X}_{1} - t \delta \hat{X}_{2}
\end{equation}

$r$ and $t$ being adjustable amplitude reflection and transmission
coefficients. If the minimum noise on this beam is below the
standard quantum limit, we are sure that the initial correlation
can only be described in a full quantum frame. We will name by the
neologism "gemellity" ("twinship") the minimum variance of this
quantity, labelled $G$, which can be found to be :

\begin{equation}
G = \frac{F_1+ F_2}{2}-\sqrt{C_{12}^2 F_1 F_2
+\left(\frac{F_1-F_2}{2}\right)^2} \label{gem}
\end{equation}
and state  : \\
\begin{displaymath}
\begin{array}{c}
 G<1 \\
\Downarrow \vspace{5pt} \\
 \textrm{\bf Impossibility of a classical description
of correlated beams}
 \end{array}
 \end{displaymath}

\subsection{Balanced case}

Let us first consider the case of two beams of equal means and
noises, so that $F_1=F_2=F$. In this case, the gemellity has a
very simple expression :

\begin{equation}
G = F (1 - |C_{12}|)
\end{equation}
The reflection and transmission amplitude coefficients $r$ and $t$
are both equal in this case to $1/\sqrt{2}$, so that $G$ can also
be written as :
\begin{equation}
G=\frac{\langle \left(\delta \hat{X}_1 - \delta \hat{X}_2\right)^2
\rangle}{2}
\end{equation}

It is nothing else than the normalized noise on the difference
between the fluctuations of the two measurements, and can be
easily monitored by simple electronic means. In the classical case
described in the previous paragraph, it is easy to show that $G$
takes the value 1, whatever the initial Fano factor $F_{in}$ of
the beam. If the gemellity G has a value smaller than 1, the two
beams have identical mean values and almost identical fluctuations
(within the quantum noise). Such beams are usually named "twin
beams", in a way reminiscent of the "twin photons" studied in the
photon counting regime. One can distinguish between "intensity
twin beams", where the measured quadrature is the amplitude
quadrature (in that case, the measured gemellity is equivalent to
the normalized difference of the intensity fluctuations of the two
beams), and which are produced by above threshold OPOs \cite{Nous,
Kumar} or by the mixing on a $50\%$ beamsplitter of a coherent
state and a squeezed vacuum \cite{Slusher}, and "quadrature twin
beams", which are produced by non degenerate OPOs below threshold
\cite{Kimble, Laurat}. The smallest measured value of the
gemellity $G$ is to the best of our knowledge $G=0.11$
\cite{Laurat2}.

The non-classical region ($G<1$) corresponds to correlations
$C_{12}$ larger than $1-1/F$. The correlation likely to produce
non-classical twin beams has a lower limit which is more and more
close to 1 when the two fields have more and more excess noise. If
each field is at shot noise, any non-zero correlation is a proof
of gemellity, and therefore of non-classical character.

\subsection{Unbalanced case}

Unbalanced beams may have also strong, or even perfect, classical
correlations. To produce classically correlated fields of unequal
intensities and fluctuations, one can use a non equal
beam-splitter with different amplitude transmission and reflection
coefficients. In this case, the correlation $C_{12}$ is found to
be
\begin{equation}
\left[C_{12}\right]_{class}=\sqrt{
(1-\frac{1}{F_1})(1-\frac{1}{F_2})} \label{correl}
\end{equation}
which is the generalization of relation (\ref{C12}). This amount
of correlation, as expected, gives a value larger than or equal to
1 to the gemellity $G$, defined by Eq(\ref{gem}).

If $F_1$ or $F_2$ is equal to 1, Expression(\ref{gem}) implies
that any non-zero correlation $C_{12}$ gives a value of G smaller
than 1 : any correlation between a field at shot noise and another
field has thus a quantum origin.

In order to experimentally determine the gemellity, one uses the
operational definition : it is the minimum noise -normalized to
shot noise - obtained when one mixes the two considered beams on a
beamsplitter of variable transmission and reflection. The
gemellity criterion for a non-classical correlation between
unbalanced beams is interesting from an experimental point of
view, because in a given experimental situation the two measured
beams do not have necessarily the same mean power and noise
\cite{Marcelo, autre,autre2}.

\section{Non-classical character of the measurement
provided by the correlation : "QND-correlated beams"}

When two observables $M_1$ and $M_2$ are perfectly correlated, the
measurement of $M_2$ gives without uncertainty the value of $M_1$.
The first measurement is thus a \textit{Quantum Non Demolition
measurement} (QND) of the observable $M_1$ performed on the second
sub-system.

We can now define a second level in the quantum character of
correlations :\\

\textbf{Quantum correlation, level 2 :} \textit{ The information
extracted from the measurement on one field provides a Quantum Non
Demolition measurement of the other.}\\

In the last decade, many studies have been devoted to the precise
definition of QND criteria\cite{QND1}, that we can use now in our
discussion. In the present case, the "Non Demolition" part of the
measurement is automatically ensured, as the measurement,
performed on beam 2, does not physically affect the measured
system, which is beam 1. Its quantum character is effective when
the measurement is able to provide enough information on the
instantaneous quantum fluctuations of the other beam so that it is
possible, using the information acquired on mode 2, to correct
mode 1 from its quantum fluctuations and transform it into a
non-classical state in the meaning of the previous section by a
feed-back or feed-forward electronic device. This criterion is
well known in QND studies\cite{QND2}, where it is shown that it is
equivalent to state that the \textit{conditional variance}
$V_{1|2}$ of beam 1 knowing beam 2 takes a value smaller than 1.
The conditional variance has the following expression in terms of
the Fano factor of beam 1 and the normalized correlation $C_{12}$
between the two :

\begin{equation}
V_{1|2}=F_1(1-C_{12}^2)
\end{equation}

\subsection{Balanced case}

Let us first consider the case where the two beams have identical
mean values and fluctuations ($F_1=F_2=F$). In this case there is
only one conditional variance $V_{1|2}=V_{2|1}=V$, and the
criterion for "QND-correlated beams" is :

\begin{equation}
V_{1|2}=V_{2|1}=V<1
\end{equation}
The conditional variance and the gemellity are related by :
\begin{equation}
V=G(1+C_{12})=2G-\frac{G^2}{F}
\end{equation}
so that :
\begin{equation}
G\leq V \leq2 G
\end{equation}

One notices that the conditional variance is always bigger than
the gemellity, so that all the QND-correlated beams are twin
beams, whereas the reverse is not true. We see also that a small
enough gemellity, namely smaller than 0.5, implies that the beams
are QND-correlated.

It is possible to show \cite{QND2} that the conditional variance
can be directly measured by using an adjustable amplification on
one of the two photocurrents, i.e. by measuring the quantity :
\begin{equation}
\hat{X}_{g}= \hat{X}_1 -g \hat{X_2}
\end{equation}
The conditional variance is equal to the minimum value of $<\delta
\hat{X}_{g}^2>$ when $g$ is varied.

\subsection{Unbalanced case}

In this case the two conditional variances are different, and
there are two possible criteria $V_{1|2} <1$ and $V_{2|1}<1$. They
are not always simultaneously satisfied : there exist situations
where $V_{1|2} <1$ and $V_{2|1}>1$ for example. This shows that
the QND criterion evaluates the correlation from the point of view
of one beam, and the information that one can have on this beam
from measurements on another one, and does not intrinsically
quantize the quantum correlation between the two fields.

It is easy to show that it is enough to have one of the two
conditional variances smaller than 1 to have twin beams. In
contrast, there are regions of the parameter space where G is
smaller than 0.5 and where one of the two conditional variances is
bigger than 1.

We will therefore give an "asymmetrical" criterion to characterize this second level of quantum correlation  : \\
\begin{displaymath}
\begin{array}{c}
 V_{1|2}<1 \quad \textbf{or}
 \quad C_{12}>\sqrt{1-\frac{1}{F_1}} \vspace{5pt}\\
\Downarrow \vspace{5pt} \\
\textrm{\bf Possibility of a QND measurement of beam 1 using the
correlation between beams 1 and 2} \\
 \end{array}
\end{displaymath}

\section{Impossibility of description by a statistical mixture of factorizable states : "inseparable beams"}

We now define a new level in the quantum character of
correlations, related to the entangled character of the state, as already stated in the introduction :\\

\textbf{Quantum correlation, level 3 : } \textit{The correlation
arises from a system which can be described only by an entangled
or non-separable quantum state.}\\

Let us first consider the case of a pure state, which is described
by a state vector $|\Psi>$. If this vector can be written as a
tensor product of states belonging to each Hilbert sub-space (i.e
it is not entangled nor non-separable), the mean value of a
product of observables $\hat{O}_1$ and $\hat{O}_2$ acting
separately in the two sub-spaces (1) and (2) will be the product
of the mean value of each observable : there will be therefore no
correlations in such a system, whatever the two observables. So,
if one finds a non-zero correlation on a single couple of
observables, even when this correlation is weak, it is a proof
that the system is in an entangled state : correlation implies
entanglement for pure cases.

Reciprocally, if the system is described by an entangled state
$|\Psi>$ of the form (\ref{entangled}), what are the conditions to
get a non-zero correlation between two observables $\hat{O}_1$ and
$\hat{O}_2$ ? One knows that $|\Psi>$ can be written in the
following form (Schmidt decomposition \cite{Houches}) :
\begin{equation}
|\Psi> = \sum_j |\psi_{1,j}> \otimes|\psi_{2,j}> \label{schmidt}
\end{equation}
where the states $|\Psi_{i,j}>$ belong to the Hilbert space of the
sub-system labelled (i) ($i=1,2$). Non-zero correlations will
happen when, firstly, the measurement on sub-system (1) is
performed on an observable $\hat{O}_1$ which has not all the
states $|\Psi_{1,j}>$ in the same eigenspace, so that the state
reduction due to the measurement changes the total state $|\Psi>$.
Secondly, in order to affect the measurement performed on an
observable $\hat{O}_2$ on system (2) the states $|\Psi_{2,j}>$
must not be in the same eigenspace of $\hat{O}_2$.

These arguments prove that the presence of entanglement in a pure
state does not imply that any couple of observable will be
correlated, and, if a correlation between two observables is
obtained, it does not imply that the correlation has reached even
the level 1 of quantum correlation. The requirement of the quantum
description of the correlation (twin beams) is therefore stronger
than the requirement of having an entangled state.

The situation is completely different if one allows the system to
be in a \textit{statistical mixture of quantum states}, so that it
is described by a density matrix instead of a state vector: in
this case, the existence of a correlation between two measured
quantities does not imply that the system is in an entangled
state. A single correlation, even perfect, between a given
observable of sub-system (1) and a given observable of sub-system
(2) can be obtained with "separable states" in the meaning of
\cite{Duan}, i.e. with states that are classical statistical
mixtures of factorizable states. They can be written as :
\begin{equation}
\rho = \sum_j p_j |\psi_{1,j}>\otimes|\psi_{2,j}>
<\psi_{1,j}|\otimes <\psi_{2,j}| \label{rhoduan}
\end{equation}
with $p_j$ positive real numbers such that $\sum_j p_j =1$. States
which cannot be written as (\ref{rhoduan}) will be called
non-separable. They are also named "entangled states" in an
extended meaning.

Let us consider as an example the system described by the
separable density matrix :
\begin{equation}
\rho = \sum_n p_n | n :1, n:2 > <n :1, n:2| \label{twinseparable}
\end{equation}
where $| n :1, n:2 >$ is a Fock state with the same number $n$ of
photons in the two modes (1) and (2). This state yields a perfect
intensity correlation between the two modes, and satisfies the two
previous criteria : the correlation $C_{12}$ is 1, and therefore
the gemellity $G$ is zero, as well as the conditional variances
$V_{1|2}$ and $V_{2|1}$.

Note that the state described by (\ref{twinseparable}) is indeed
very "quantum", in spite of not being entangled or non-separable,
as it is built from Fock states having exactly the same number of
photons in the two modes, which cannot be produced classically,
but only through cascade processes, such as parametric
down-conversion. We see here that quantum correlations and
entanglement are different notions, which are of course related,
but not in a straightforward and simple way.

Duan et al. \cite{Duan} have shown that in order to ascertain the
separable character of the physical state of a system, one needs
to make \textit{two joint correlation measurements on
non-commuting observables} on the system, and not only one, as was
the case in the two previous sections. They have shown that in the
case of Gaussian states, there exists a necessary and sufficient
criterion of separability in terms of the quantity $S_{12}$, that
we will call "separability", and is given by :
\begin{equation}
S_{12}=\frac{1}{2}\left(\langle
\delta(\hat{X}_{+1}-\hat{X}_{+2})^2\rangle+\langle
\delta(\hat{X}_{-1}+\hat{X}_{-2})^2\rangle \right)
\end{equation}
The separability $S_{12}$ appears as the sum of the gemellity
$G_+$ measuring the correlations between real quadrature
components of the two beams, and the (anti)gemellity $G_-$
measuring the anticorrelation between the imaginary quadrature
component of the same beams (defined with a $+$ instead of a $-$
in equation (\ref{dif})).

The third level of quantum correlation is evaluated by the well-known Duan criterion, which writes : \\
\begin{displaymath}
\begin{array}{c}
S_{12}<2 \vspace{5pt}\\
\Downarrow \vspace{5pt} \\
\textrm{\bf Quantum correlation arising from an entangled or
non-separable state}
\end{array}
\end{displaymath}

This criterion allows us to establish some relations between the
different levels of quantum correlations that we have already
considered.

For example, classical beams will give values larger than 1 for
the gemellities measured on any variables, and in particular on
$\hat{X}_+$ and $\hat{X}_-$. In this case, the quantity $S_{12}$
is larger than 2, and the two beams are therefore separable. A
contrario, non separable beams imply that at least one of the two
gemellities is smaller than 1, and therefore that the beams are at
least "twins", in intensity or in phase. For quadrature
measurements on statistical mixtures of Gaussian states, the
non-separability criterion implies that the criterion 1 is
fulfilled and is therefore stronger than this latter one. Note
that the beams are not necessarily QND-correlated in one of these
variables., so that level 2 is not necessarily reached.

Non separable beams are usually prepared by mixing two
non-classical states, such as squeezed states, on a beamsplitter
\cite{mixing}, but it has been shown \cite{Vincent} that one can
generate an entangled state from a single squeezed beam mixed with
a coherent state plus some well adapted linear processing of the
two output beams.

\section{Possibility of QND measurement of two conjugate variables : "EPR beams"}

In their famous paper, Einstein, Podolsky and Rosen \cite{EPR}
have exhibited the following wavefunction for the continuous
variables position $\hat{x}_i$ and momentum $\hat{p}_i$ ($i=1,2$)
of two particles :
\begin{equation}
\psi(x_1,x_2)=\int_{-\infty}^{+\infty} e^{ip(x_1-x_2+x_0)/\hbar}dp
\end{equation}
where $x_0$ is a constant, and shown that it provides perfectly
correlated position measurements and perfectly anti-correlated
momentum measurements of the two particles. This state, which is
obviously entangled, can be readily transferred in the domain of
quadrature operators of two light modes \cite{Reid1}. In quantum
optics terms, it allows us to perform perfect QND measurements of
the two quadrature components of mode 1, by measurements performed
only on beam 2. This perfect information that one eventually gets
on the two quadratures of the field apparently contradicts the
fact that these measurements are associated to two non-commuting
operators, and therefore obey a Heisenberg inequality.

We now reach a fourth level in the quantum character
of the correlations :\\

\textbf{Quantum correlation, level 4 : } \textit{The information
extracted from the measurement of the two quadratures of one field
provides values for the quadratures of the other which violate the
Heisenberg inequality}\\

This situation has been extensively considered and discussed by M.
Reid and co-workers \cite{Reid1, Reid2}, which have shown that
this violation is only apparent, and does not violate the basic
postulate of Quantum Mechanics. They have introduced the
following criterion to characterize this fourth level of quantum correlation of the so-called "EPR beams"  : \\
\begin{displaymath}
\begin{array}{c}
 V_{X_{+1}|X_{+2}}V_{X_{-1}|X_{-2}}< 1 \vspace{5pt} \\
 \Downarrow \vspace{5pt} \\
\textrm{\bf Possibility of an apparent violation of the Heisenberg
inequality}\\
\textrm{\bf for the quadratures components of beam 1 through
measurements performed on beam 2}
\end{array}
\end{displaymath}
where $V_{X_{+1}|X_{+2}}$ is the conditional variance of $X_{+1}$
knowing $X_{+2}$, and $V_{X_{-1}|X_{-2}}$ is the conditional
variance of $X_{-1}$ knowing $X_{-2}$.

This condition is related somehow to the QND-correlated beams of
section 4. It can be written in terms of the normalized
correlation $C_{X_{+1}X_{+2}}$ and anticorrelation
$C_{X_{-1}X_{-2}}$ :
\begin{equation}
\left(1- C_{X_{+1}X_{+2}}^2\right)\left(1-
C_{X_{-1}X_{-2}}^2\right)> \frac{1}{F_+ F_-}
 \end{equation}
where $F_+$ and $F_-$ are the noise variance on quadratures $+$
and $-$ normalized to shot noise (fulfilling $F_+F_->1$).

The EPR correlation turns out to be stronger than the
non-separability correlation, in the same way as the QND criterion
of section 4 is stronger than the non-classical criterion of
section 3 : it has been shown \cite{Bowen} that all EPR beams are
non-separable, whereas the reverse is not true. In the same
article, {\it Bowen et al.} show that for pure states the two
criteria correspond to the same physical states. However, {\it
Duan} criteria is robust with respect to the mixed character of
the fields states, whereas the EPR criteria can not be fulfilled
in the presence of more than 50$\%$ of losses. Let us stress that
this behaviour is linked to the {\it no-cloning} theorem : it has
been proved that linear amplification and a 50/50 beam-splitter
produces the best possibles two copies (clones) of any input state
\cite{Cloning}. Hence, when two beams are EPR correlated, i.e.
that less than 50{\%} losses are present, on a quantum information
point of view we are sure than no spy has a better copy of the
state. This is in the same way relevant for the success of
teleportation \cite{Grangier}.

\section{Impossibility of description of the multiple correlation by local classical stochastic variables : "Bell beams"}

Quantum fluctuations can be mimicked in many instances by
classically-looking stochastic supplementary variables. This is
the case in particular when one uses the approach of "vacuum
fluctuations", behaving as classical fluctuations, but with a
variance given by quantum mechanics and carrying no energy. This
is to be distinguished from the classical stochastic fields,
originating from the uncontrolled variations of the classical
parameters of the light source. Bell \cite{Bellineq} has shown
that such a stochastic modelling was not likely to reproduce all
the correlations that can be encountered in quantum mechanical
systems when these supplementary stochastic variables (usually
named "hidden variables") were local, i.e. attached to the
sub-system under measurement. He introduced inequalities fulfilled
by any local hidden variable models, and violated in some very
specific situations of quantum mechanical correlated systems.

We must therefore introduce a new level of quantum correlations
:\\

\textbf{Quantum correlation, level 5 : } \textit{The multiple
correlations of the system cannot be described by local hidden
variable approaches}\\

The corresponding criterion for this level of quantum correlation
is the celebrated Bell inequality \cite{Bellineq}. We will not go
into the details of it here for the following reason : one shows
in Quantum Optics that all phenomena can be described through the
use of quasi-probability distributions \cite{Walls, Bachor}, such
as the Wigner representation or others. For the special case of
light beams having a Gaussian statistics, which the case of all
the physical situations encountered so far in the regime of
continuous variables in optics, the Wigner representation is
everywhere positive : the quasi-probability distribution becomes a
true probability distribution, the evolution of which can be
mapped into stochastic equations for fluctuating fields : these
stochastic fields constitute in this case the local "hidden"
variables which account for all the observed quantity, including
the variances and the correlations between measurements. This
means that there is never a violation of the Bell inequality in
the continuous variable regime with Gaussian states, and the level
5 of quantum correlations is never reached in this case.

To reach it, one needs to deal with non Gaussian states, with
partly negative Wigner functions, such as the Fock states,
Schr\"{o}dinger cat states \cite{Cat}, or states produced through
conditional non-Gaussian measurements like photon-counting. The
discussion of such situations is beyond the scope of this simple
introductory paper on quantum correlations.

\section{Conclusion}

The exploration of the quantum world, in which professor Walther
has undoubtedly played a major role, has unveiled physical
situations which are looking more and more strange for an observer
only acquainted to the certainties of classical physics. We have
tried in this short review to assess and classify the "degree of
oddness" of quantum optical phenomena. In the last decades,
theoreticians and experimentalists have gone higher and higher in
such a ladder of pure quantum effects. There is no doubt that they
are far from reaching the top of the quest, and that new heights
of even stranger quantum properties will be attained in the
future.

\section{Acknowledgments}

We would like to thank Philippe Grangier and Thomas Coudreau for
fruitful discussions. Laboratoire Kastler Brossel, of the Ecole
Normale Sup\'{e}rieure and University Pierre et Marie Curie, is
associated to CNRS. This work has been supported by the European
Union in the frame of the QUANTIM network (contract IST
2000-26019).

\end{document}